%% file: body.tex
 \documentclass[manuscript]{acmart}  

\usepackage{booktabs} 
\usepackage{url}  
\usepackage[ruled]{algorithm2e} 
\usepackage{array}
\usepackage{xpatch}
\newcolumntype{M}[1]{>{\centering\arraybackslash}m{#1}}

\begin{document}

\settopmatter{printacmref=false} 
\renewcommand\footnotetextcopyrightpermission[1]{} 
\pagestyle{plain} 
\setcopyright{none}
\makeatletter
\xpatchcmd{\ps@firstpagestyle}{Manuscript submitted to ACM}{}{\typeout{First patch succeeded}}{\typeout{first patch failed}}
\xpatchcmd{\ps@standardpagestyle}{Manuscript submitted to ACM}{}{\typeout{Second patch succeeded}}{\typeout{Second patch failed}}    \@ACM@manuscriptfalse
\makeatother


\title{Audio-Based Activities of Daily Living (ADL) Recognition with Large-Scale Acoustic Embeddings from Online Videos}

\author{Dawei Liang}
\affiliation{
  \institution{the University of Texas at Austin}
  \country{USA}}
\email{dawei.liang@utexas.edu}

\author{Edison Thomaz}
\affiliation{
  \institution{the University of Texas at Austin}
  \country{USA}}
\email{ethomaz@utexas.edu}


\begin{abstract}
Over the years, activity sensing and recognition has been shown to play a key enabling role in a wide range of applications, from sustainability and human-computer interaction to health care. While many recognition tasks have traditionally employed inertial sensors, acoustic-based methods offer the benefit of capturing rich contextual information, which can be useful when discriminating complex activities. Given the emergence of deep learning techniques and leveraging new, large-scaled multi-media datasets, this paper revisits the opportunity of training audio-based classifiers without the onerous and time-consuming task of annotating audio data. We propose a framework for audio-based activity recognition that makes use of millions of embedding features from public online video sound clips. Based on the combination of oversampling and deep learning approaches, our framework does not require further feature processing or outliers filtering as in prior work. We evaluated our approach in the context of Activities of Daily Living (ADL) by recognizing 15 everyday activities with 14 participants in their own homes, achieving 64.2\% and 83.6\% averaged within-subject accuracy in terms of top-1 and top-3 classification respectively. Individual class performance was also examined in the paper to further study the co-occurrence characteristics of the activities and the robustness of the framework.
\end{abstract}

%
%

\ccsdesc[500]{Human-centered computing~Empirical studies in ubiquitous and mobile computing}

%
%

\keywords{Activity recognition, deep learning, multi-class classification,
audio processing}

\maketitle

\input{introduction}

\input{related-work}

\input{vggish}
\input{implementation}

\input{feasibility-study}

\input{Free_test}

\input{conclusion}

\bibliographystyle{unsrt}
\bibliography{reference}

\end{document}

%% file: introduction.tex
\section{Introduction}

Sensing and recognizing human daily activities has been demonstrated to be useful in many areas, from sustainability to health care. For example, older adults in their own homes could benefit from proactive assistance and monitoring in as a way to "live-in-place" and not be forced to move to an assisted-living or nursing facility. While on-body inertial sensors such as accelerometers and gyroscopes are popular in many human activity recognition applications, prior work suggests that they are not effective at recognizing complex and multidimensional activities on their own \cite{ravi2005activity,kwapisz2011activity,anjum2013activity}. Audio, on the other hand, offers much promise in this respect; many daily activities generate characteristic sounds that can be captured with any off-the-shelf device with a microphone. Hence, researchers have proposed several different types of audio event recognition frameworks over the years, from applications on wearable and mobile devices \cite{thomaz2015inferring,rossi2013ambientsense} to home-based sensor systems \cite{laput2017synthetic,chen2005bathroom}. 

With the development of deep neural networks in recent years, several efforts have been made by researchers to model large-scaled acoustic events. These include the usage of deep learning for sound classification on existing datasets \cite{salamon2017deep} and the recognition of acoustic categories in the wild \cite{lane2015deepear}. However, most such frameworks suffer from the laborious collection of ground truth training data. Some researchers have explored the use of crowd-sourced data to alleviate the problem, such as Nguyen et al. and Rossi et al. Despite encouraging results, these methods have proven difficult to scale as they partially rely on human input or interaction. 

Large-scale, open-source audio collections now offer a rich source of audio data reflecting a large number of everyday activities. In this work, we present a novel scheme to recognize activities of daily living in the home. Instead of directly collecting ground truth data and labels from users as in most prior research, we explored the feasibility of using millions of audio embeddings from general-sourced YouTube videos as the only training set. Due to the considerable size and highly unbalanced characteristics of the on-line data, our method combines both oversampling and deep learning approaches. The contributions of this work can be summarized as:

\begin{itemize}
  \item A novel framework for activity recognition with ambient audio that relies exclusively on a large-scale audio dataset. It aims to empower traditional audio-based activity recognition by applying over 2 million audio embedding features from nearly 52,000 public Youtube videos. Typically, the proposed framework does not require feature augmentation and semi-supervised learning processes as in relevant research.
  \item An evaluation of the framework with 14 subjects in their homes and 15 activities of daily living. The proposed method was able to yield promising performance for activities and was robust to environmental variability.
\end{itemize}

%% file: related-work.tex
\section{Related Work}

\subsection{Inertial Sensing}
Activity recognition based on sensor data is not new. It has been widely used for several domains including self monitoring \cite{thomaz2015practical}, assistance in smart home \cite{chen2012knowledge} or diagnosis of some activity-related disease \cite{grunerbl2015smartphone}. Traditional approaches of activity recognition rely on inertial sensors such as accelerometers and gyroscopes. They can be implemented flexibly on smart phones \cite{kwapisz2011activity,anjum2013activity}, smart watches \cite{thomaz2015practical,shoaib2015towards} or wearable sensor boards \cite{ravi2005activity}. For example, Kwapisz et al. \cite{kwapisz2011activity} managed to recognize walking, jogging, upstairs, downstairs, sitting and standing by just using a smart phone in subjects' pockets. Thomaz et al.\cite{thomaz2015practical} proposed the usage of 3-axis accelerometers embedded in an off-the-shelf smart watch for detection of eating moment. Similarly, Ravi et al. \cite{ravi2005activity} showed the feasibility of attaching a sensor board on human body for simple movement classification. Most of such work focused on recognition of very simple activities involving limited types of sensors. The limitation of the activity class availability can be improved by combining more sensing modalities. However, implementation of complex sensor arrays always brings challenges in insufficiency of training sets and constraints of energy or computing resource. The subject and location sensitivity of traditional inertial sensing can make it even harder for generalization of activity models \cite{su2014activity}. 

\subsection{Audio Sensing}
Out of such reasons, researchers have proposed to empower activity recognition by video and audio approaches. Microphones have the benefits of simplicity and flexibility for implementation. Eronen et al. \cite{eronen2006audio} proposed the pilot study to recognize common contexts based on sounds by using statistical learning methods. Yatani and Truong \cite{yatani2012bodyscope} explored the recognition of 12 activities related to throat movement such as eating, drinking, speaking and coughing by acoustic data collected from human throat area. This was completed by a simple wearable headset consisting of a tiny microphone, a piece of stethoscope and a Bluetooth module. Another study showed that human eating activity can also be effectively inferred by using wrist-mounted acoustic sensing \cite{thomaz2015inferring}. This implies the practicality of simple audio-based activity recognition by off-the-shelf products such as smart watches. With the development of smart phones in recent years, phone-based acoustic sensing also shows great capability on activity recognition tasks. The \textit{AmbientSense} application \cite{rossi2013ambientsense} is an example. It is an Android app that can process ambient sound data in real time either on user front end or on an on-line server. It was tested on mainstream smart phones (Samsung Galaxy SII and Google Nexus One) and yielded satisfactory results on classification of 23 context of daily life. Lu et al. \cite{lu2009soundsense} developed the \textit{SoundSense} to detect multiple speech, music and ambient sound categories based on mobile platforms. Acoustic sensing can also be used for indoor scenarios, especially when video-based methods may bring privacy concerns. Laput et al.\cite{laput2017synthetic} described the concept of general-purpose sensing, where multiple sensor units including a microphone were embedded on a single home-oriented sensor tag. Chen et al. \cite{chen2005bathroom} provided an audio solution for detection of 6 common activities in bathroom based on MFCC features. Their work typically aims at elder care since direct behavioral observations can be quite embarrassing to be shared with clinicians. More recently, acoustic sensing and recognition have been significantly improved based on the usage of deep learning techniques. Salamon and Bello \cite{salamon2017deep} proposed an architecture combining feature augmentation and a CNN to evaluate on-line audio data. Lane et al. \cite{lane2015deepear} developed the \textit{DeepEar} to classify multiple categories for different sensing tasks based on a well-tuned fully connected network.

\subsection{Audio-Based Activity Recognition with Online Data}
Most of the prior work requires a manual collection of ground truth audio data from individual users. This can be quite laborious especially if we are targeting at multiple classes of activities. Also, it is actually unrealistic to ask the users to train the model on their own before using it. Hence, Hwang and Lee \cite{hwang2012environmental} introduced a crowd-sourcing framework for the problem. They developed a mobile platform to collect audio data from multiple users. The platform could then generate a global K-nearest neighbors (KNN) classifier based on Gaussian histogram of MFCC features to recognize basic audio scenes. However, this still requires collection of user data and the performance of the system highly depends on the size and quality of the training set. General-purposed acoustic database, on the other hand, can potentially serve as ideal data source to the existing systems. Over the past years, the \textit{Freesound} database (\url{https://freesound.org/}) has been one of the most commonly used database for audio research. Started in 2005 and currently maintained by the Freesound team, it is a crowd-sourced dataset consisting of over 120'000 annotated audio recordings. Besides, Salamon et al. \cite{salamon2014dataset} released the \textit{UrbanSound} database containing 18.5 hours of urban sound clips selected from the Freesound. S\"ager et al. \cite{sager2018audiopairbank} improved the Freesound recordings by adding adjective-noun and verb-noun pairs to the audio tags and constructed a new \textit{AudioPairBank} dataset. Rossi et al. \cite{rossi2012recognizing} first attempted context recognition based on MFCC features extracted from the on-line Freesound database by using a Gaussian Mixture Model (GMM). However, due to the limited size of the training set (4678 audio samples for 23 target context), the top-1 classification accuracy based on dedicated sound recordings was just 38\%. The performance was improved to 57\% by manually filtering over one third of the samples as outliers. Beyond that, Nguyen et al. \cite{nguyen2013combining, nguyen2013towards} leveraged semi-supervised learning methods to combine the on-line Freesound data with users' own recordings. After manually filtering outliers for quality, they trained a semi-supervised GMM on MFCC features extracted from 163 Freesound audio clips for 9 context classes. The model was then applied to unlabeled user-centric data recorded by smart phones with a headset microphone. The performance was evaluated based on the second half of the user data with an average accuracy of 54\% for 7 users. To further improve the performance, Nguyen et al. \cite{nguyen2013towards} also presented two active learning mechanisms, where a supervised GMM was first trained on the same Freesound data or well-labeled user data and then interactively queried users for labeling the unlabeled user-centric data. Clearly, from the prior work we can see that the existing crowd-sourced datasets do not generalize sufficiently the audio recorded across users, and previous research still needs to rely on user data and manual filtering of outliers for better performance.

In April 2017, Google's Sound Understanding team released the \textit{Audio Set} database \cite{gemmeke2017audio}\footnote{\url{https://research.google.com/audioset/}} containing ontology and embedding features of over 2 million audio clips drawn from general YouTube videos. The clips are categorized by 527 audio labels and many of the class labels can be potentially bridged to the common real world scenarios. Actually, the idea of domain adaptation from the web has been developed in several activity recognition research. Hu et al. \cite{hu2011cross} proposed to use web search text as a bridge for similarity measures between sensor readings. Fast et al. \cite{fast2016augur} developed the \textit{Augur}, a system leveraging contexts from on-line fictions to predict human activities in the real world. In terms of audio-based classification, Aytar et al. \cite{aytar2016soundnet} described the \textit{SoundNet} framework for knowledge transfer between large-scaled videos and target sounds based on a deep CNN. To the best of our knowledge, however, very few attempts have been made to adapt such tremendous scale of on-line audio samples for real-world activity recognition, and this can be even challenging when leveraging the YouTube sound features due to the ambiguous source of the raw videos from movies, cartoons to crowd-sourced data. The most relevant up-to-date achievement was proposed by Laput et al. \cite{laput2018ubicoustics}, where the researchers  developed a mixed process of audio augmentation for a deep network and combined the online sound effect libraries with the Audio Set data for audio context classification. Their work shows promising results when applying the augmentation process with the online sound effect data. However, the performance of the framework dropped significantly if purely using the video sounds (i.e. the Audio Set \cite{gemmeke2017audio} data) without augmentation. Moreover, their work mainly focused on the classification of environmental contexts, and the statistics in terms of individual activity classes still largely remained unexplored. In our research, we aimed to study the feasibility and performance reported from the perspective of individual activity recognition by leveraging only the online video sound clips for training. Our in-lab and multi-subject studies showed that the proposed framework was able to yield promising performance even without any feature augmentation or semi-supervised learning techniques.

%% file: implementation.tex
\section{Implementation}
\subsection{\textit{Audio Set}}

In 2017, Google's Sound Understanding team released a large-scale acoustic dataset, named Audio Set \cite{gemmeke2017audio}, endeavoring to bridge the gap in data availability between image and audio research. The Audio Set contains information of over 2 million audio soundtracks drawn from general YouTube videos. The dataset is structured as a hierarchical ontology consisting of 527 class labels and the size is still growing now. All audio clips are equally chunked as 10 seconds long and labeled by human experts. 

The dataset does not provide original waveforms of the audio clips. Instead, the samples are presented in the form of both source indexes and bottleneck embedding features. The audio index contains information of the audio ID, URL, class labels, and start and end time of the sample within the corresponding source video. The embedding features are generated from the embedding layer of a VGG-like deep neural network (DNN) architecture trained on the YouTube-100M dataset \cite{hershey2017cnn}. The generation frequency is roughly 1Hz (96 10ms audio frames, i.e. 0.96 seconds of audio per embedding vector). In other words, one embedding vector can describe one second of audio clip, and therefore there are 10 embedding vectors for each audio clip within the dataset. Before released, the embedding vectors have also been post-processed by principle component analysis (PCA) and whitening as well as quantization to 8 bits per embedding element. Only the first 128 PCA coefficients are kept and released.

The original vectors are all stored within TensorFlow \cite{tensorflow2015-whitepaper} Record files. Given the significant size of the embeddings and the lack of convenience for data processing, Kong et al. \cite{kong2017audio} provided a converted Python Numpy version of the raw embeddings which are adopted in our research. Their converted dataset has been released publicly online\footnote{\url{https://github.com/qiuqiangkong/ICASSP2018\_audioset}}.

\subsection{Label Association}

Before implementation, we need to consider the range of target activities and how we can associate the class labels in the Audio Set with them. Our research leverages existing audio samples and labels from on line as the training set, and we aim at target activities that frequently appear in the home. Specifically, the range of our target classes has been limited to target activities that are suitable for audio-based recognition. Here `suitable' means that the sound of the activity should be featured and easily captured in practice. Hence, we excluded pet categories for our study because such sound is normally sparse in natural home scenarios. Classes such as 'silence' was also not chosen because the corresponding attributes can be ambiguous from sleeping, standing, to maybe just absence of the person in the room. Body movement with very weak sound features is not suitable for audio-based recognition as well. Further more, it is not always possible to find an exact matching between the Audio Set labels and the actual activities. In such cases, we adopted an indirect matching process. That is, we first determined the most relevant objects and environmental contexts regarding to the target activities. We then chose Audio Set classes of such objects and contexts as representation of the activities. For example, we used class 'water tap' and 'sink' as representation of 'washing hands and faces' as all three classes involve usage of water and the features are quite similar. This is actually a very subjective process as there is no quantized measurement to determine the similarity between such relevant classes and the actual target classes. For the class 'listening to music', we focused on studying only piano-related musics as examples.

It is noted that the dataset provides a quality rating of audio labels based on manual assessment. Most of the labels have been assessed by experts based on a random check of 10 audio segments within the label. The samples of each label are actually divided into three subsets (\textit{evaluation}, \textit{balanced training}, and \textit{unbalanced training}) for training and evaluation purposes. The evaluation and balanced training sets are of much smaller size than the rest unbalanced training set, and due to the considerable size of samples and factors such as misinterpretation or confusibility, many class labels of the unbalanced training sets are actually of poor rating results. In our framework, we did not consider the sample ratings and we incorporate all three evaluation, balanced training and unbalanced data for our training set.

\begin{table}[t]
\caption{Target activities and association with Audio Set labels}
\label{tab:one}
\begin{minipage}{\columnwidth}
\begin{center}
\begin{tabular}{lll}
  \toprule
  \textbf{Category}   & \textbf{Activity Class}   & \textbf{Associated Audio Set Labels} \\
    & Bathing/Showering   & Bathtub (filling or washing) \\
    & Washing hands and face   & Sink (filling or washing); Water tap, faucet \\
    Bathroom & Flushing toilet   & Toilet flush \\
    & Brushing teeth   & Toothbrush \\
    & Shavering   & Electric shaver, electric razor \\ \hline
    & Chopping food   & Chopping (food) \\
    & Frying food   & Frying (food) \\
    Kitchen & Boiling water    & Boiling \\
    & Squeezing juice   & Blender \\
    & Using microwave oven   & Microwave oven \\ \hline
    & Watching TV   & Television \\
    Living/Bed room & Listening to music    & Piano \\
    & Floor cleaning   & Vacuum cleaner \\
    & Chatting   & Conversation; Narration, monologue \\ \hline
    Outdoor  & Strolling   & Walk, footsteps; Wind noise (microphone) \\
  \bottomrule
\end{tabular}
\end{center}

\end{minipage}
\end{table}%

We therefore determined 15 common home-related activities for the framework. They are associated with 18 Audio Set labels. Table 1 shows the association between our target activities and the Audio Set class labels, and all audio embeddings of the listed Audio Set classes are used as the only training data in our proposed scheme.

\subsection{Oversampling}

A typical characteristic of the Audio Set data is the unbalanced distribution in terms of the class size. In our implementation, we also removed samples with label co-occurrence among the target classes to ensure mutual exclusiveness, and table 2 shows the number of embedding vectors per class in our raw training set without any sampling process. The numbers here include embeddings from all three subsets (evaluation set, balanced training set and unbalanced training set). The actual size for some classes is slightly smaller than they appear on the released Audio Set since we adopted the converted Python Numpy version of features as mentioned. As we can see, classes 'chatting' and 'listening to music' have the most embeddings (174,220 and 115,200 respectively). Class 'brushing teeth' is of the least, only 1230, which accounts for 0.7\% of the largest class. In other words, the two majority classes account for over half of the whole training set. The unbalanced distribution of the class size leads to highly unbalanced training in our study. As we will see in the dedicated test section, the distribution of training class can heavily affect the recognition performance, and we implemented two oversampling processes for the problem.  

\begin{table}[b]%
\caption{Number of embedding vectors per activity class}
\label{tab:two}
\begin{minipage}{\columnwidth}
\begin{center}
\begin{tabular}{ll}
  \toprule
  \textbf{Activity Category}   & \textbf{\# of Embedding Vectors} \\
    Chatting   & 174,220 \\
    Listening to music    & 115,200 \\
    Strolling in courtyard   & 81,450 \\
    Watching TV   & 22,250 \\
    Flushing toilet   & 22,190 \\
    Floor cleaning   & 19,710 \\
    Washing hands and face   & 17,080 \\
    Frying food   & 15,820 \\
    Bathing/Showering   & 14,270 \\
    Squeezing juice   & 12,600 \\
    Shavering   & 8,570 \\
    Using microwave oven   & 8,180 \\
    Boiling water    & 4,440 \\
    Chopping food   & 2,060 \\
    Brushing teeth   & 1,230 \\
    Total   & 519,270 \\
  \bottomrule
\end{tabular}
\end{center}

\end{minipage}
\end{table}%

The unbalanced distribution of labels can mainly be affected by two facts. Firstly, the distribution actually reflects the diversity and frequency of the class labels within the source YouTube videos. For example, elements of chatting or musics can be captured in a large amount of video topics, from advertisement, news to cartoons. Brushing teeth, on the contrary, appears much less and typically just in some movie scenes or daily life recordings. Chatting can also involve several modalities according to the speaker's gender, age and the context of the speech, while brushing activities seem to be much more similar among each. Secondly, we are using only samples without label co-occurrence among the target classes. The size of the remaining disjoint data can also affect the actual distribution in our training set.

The effects of unbalanced training on classification have been discussed in several work \cite{liu2007generative,chawla2002smote,han2005borderline}. Without prior knowledge of the unbalanced priors, a classifier can always tend to predict the majority classes, and there should be higher cost for misclassifying the minority classes \cite{liu2007generative}. In our scheme, we implemented random oversampling with replacement and synthetic minority oversampling technique (SMOTE) \cite{chawla2002smote} to handle the problem. The process of random oversampling can be divided into two steps. The first is to calculate the sampling size for each minority class, i.e. to calculate the difference of size between the target class and the majority class. Then each minority class will be re-sampled with replacement until the sampling size is filled. This is actually replication of existing data without introducing any extra information into the dataset. The SMOTE, on the contrary, works by adding new elements for the minority classes. It leverages the K-nearest-neighbors (KNN) approach to first generate new data points around the existing data points. Then one of the neighbors is randomly selected as the synthetic new elements and will be introduced to the minority class. In our implementation, the oversampling process was developed based on the Python imbalanced-learn package \cite{JMLR:v18:16-365,chawla2002smote}. All parameters were set as default in the imbalanced-learn package version 0.3.3 except that the random state was kept as 0. By the oversampling processes, we actually obtain 2,613,300 embedding vectors in total for the 15 classes. The size are the same for both random oversampling and the SMOTE.

\subsection{Architecture}

\begin{figure}[b]
  \includegraphics[width=0.75\textwidth]{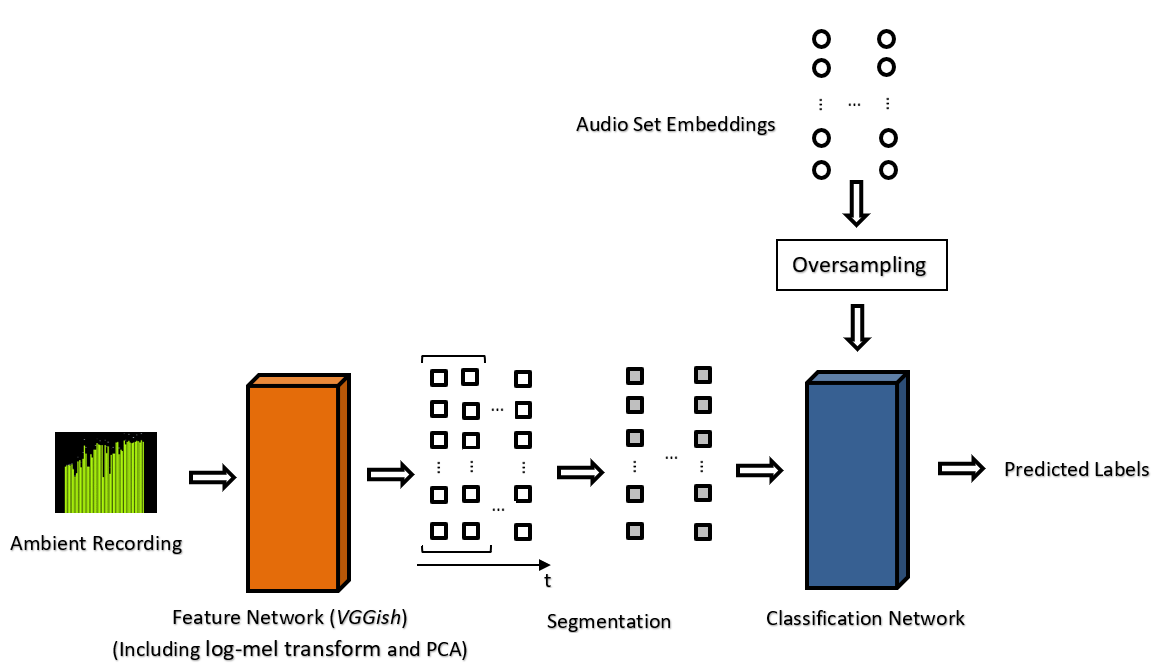}
  \caption{Architecture of our proposed scheme. We applied the VGGish model \cite{hershey2017cnn} as the feature extraction network. The feature network was pre-trained on the YouTube-100M dataset and all parameters were fixed in our training process. The generated embeddings are then segmented and passed to the classification network. Our classification network consists of plain 1-dimensional convolutional layers and dense layers, and the model was trained and fine-tuned on the oversampled Audio Set \cite{gemmeke2017audio} embeddings.}
  \label{fig:one}
\end{figure}

\begin{figure}[t]
  \includegraphics[width=0.75\textwidth]{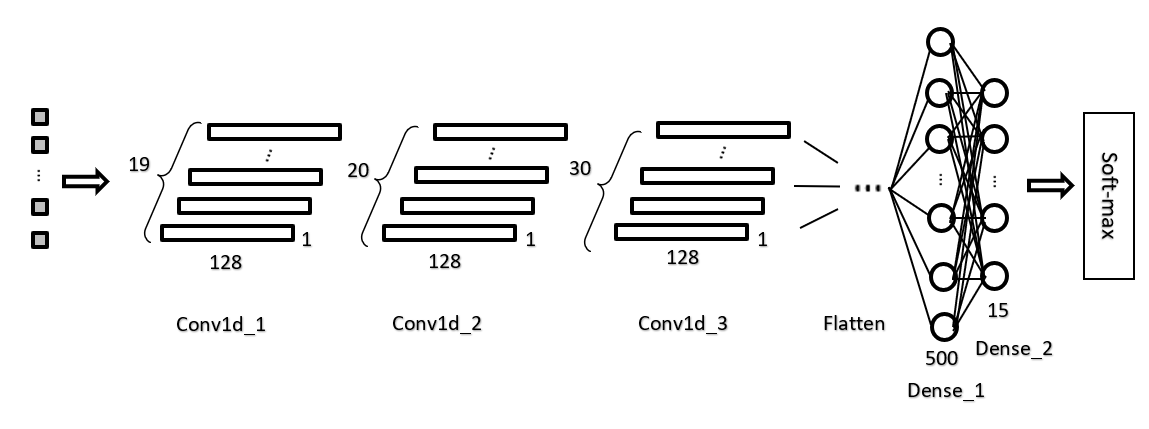}
  \caption{Architecture of the classification network. The classification network is constructed as 3 1-dimensional convolutional layers and 2 fully connect (dense) layers. The network takes as input segmented embedding vectors and outputs probability distribution of the activity labels.}
  \label{fig:two}
\end{figure}

Deep learning has been proven to be powerful for large-scale classification. Due to the considerable size of audio samples involved in our study, and also to keep the same feature format as released in the Audio Set, we adopted neural networks for both embedding feature extraction and classification in our proposed framework.
Fig.1 shows the architecture. Overall, there are two networks in our mechanism, a pre-trained feature extraction network and a classification network. In details, we adopted the pre-trained \textit{VGGish} model \cite{hershey2017cnn} as the extraction network and all parameters of the network were fixed during our training process. The classification network consists of 1-dimensional convolutional layers and dense layers. The parameters and weights of the classification network were trained and fine tuned on the Audio Set data. Besides, we added an embedding segmentation process between the two networks to improve recognition performance.

In the initial Audio Set, the frame-level features of the audio clips were generated by a VGG-like acoustic model pre-trained on the YouTube-100M dataset. To enable researchers to extract the same format of features, Hershey et al. \cite{hershey2017cnn} provided a TensorFlow version of the model called \textit{VGGish}. It has been trained on the same YouTube-100M dataset and can produce the same format of 128-dimensional embeddings for every second of audio sample. The VGGish model takes as input non-overlapping frames of log mel spectrogram of raw audio waveforms. The source codes and weights of the pre-trained VGGish model are released on the public Audio Set model GitHub repository\footnote{\url{https://github.com/tensorflow/models/tree/master/research/audioset}}. The source codes also include pre-processing steps for extracting the log mel spectrogram features to feed the model and post-processing steps for PCA transform and element-wise quantization which have also been adopted on the released Audio Set data. In our implementation, the audio pre-processing step takes as input audio waveforms with 16 bit resolution, so we manually convert other formats of audio samples (such as raw recordings from smart phones) to the wave format using a free on-line converter\footnote{\url{https://audio.online-convert.com/convert-to-wav}} before passing the raw audio for processing. The parameters of the VGGish network kept constant during the whole training and validation process. The network could then output a vector of 128 syntactic embeddings for every second of the input audio.

Our classification network consists of 3 plain convolutional layers and 2 dense (fully connected) layers. The structure is shown in Fig.2. The convolutional layers are all 1-dimensional tensor with linear activation and same paddings to ensure the same feature size. The number of channels are 19, 20 and 30 respectively for the 3 layers. The kernel size was all set as 5 with a stride of 1. We applied 500 neurons for the first dense layer. The second dense layer is the output layer, thus there are 15 neurons and the output activation was set as softmax. A flatten layer was used to connect the convolutional layers and the dense layers. We chose categorical cross entropy as the loss. In terms of the optimizer, we applied stochastic gradient descent with Nesterov momentum. The learning rate was set as 0.001 with 1e-6 decay and 0.9 momentum. The network takes as input (128 * 1) segmented and normalized embeddings from the segmentation step of our architecture and outputs predicted probability distribution of the labels. Under the top-1 classification scenario, the label with the highest probability will be selected as the final prediction. Our classification network was built and compiled on Python Keras API \cite{chollet2015keras} with Tensorflow \cite{tensorflow2015-whitepaper} backend. The weights were trained and fine tuned on the Audio Set embeddings.

In addition to the neural networks and audio processing steps, we also applied embedding segmentation to determine the unit length of an audio segment for recognition. This is natural because the length of a single embedding vector (1 second) can be too short to some activities and may not be able to capture enough information for recognition. Also, increasing the segment length can help to alleviate the effects of outliers and noise within the real world recordings. Hence, we introduced a segmentation process on embeddings between the two networks. For convenience, in the following sections we will describe the length of a unit segment by number of embedding vectors (1 second each). In our architecture, the segmentation is completed by grouping the embedding vectors using a fix-sized window with no overlaps. The vectors will then be averaged within each group to yield a new embedding vector. In other words, each unit audio segment is described by an averaged embedding vector. Activity labels will then be assigned to the averaged vectors and those vectors actually serve as the instances for classification. The embeddings are standardized using min-max scaling before fitting to the classification network.

The source code of our overall architecture has been made publicly available online\footnote{\url{https://github.com/dawei-liang/AudioAR_Research_Codes}}. Both the oversampling and training processes were developed on the Texas Advanced Computing Center (TACC) Maverick server. Specifically, we applied the NVIDIA K40 GPU on the server to accelerate the training process. The training embeddings were split as 90\% for training and 10\% for validation using the Pyhton Scikit-learn package \cite{scikit-learn}. The TensorFlow version provided was TensorFlow-GPU 1.0.0 \cite{tensorflow2015-whitepaper}. Before training, we set all random seeds as 0 to ensure the same training status. Besides, a batch of 100 embedding vectors were input each time. The classification network was trained until the validation performance no longer improved (in our study, 15 to 20 epochs depending on the re-sampling set in use).

%% file: feasibility-study.tex
\section{Feasibility Study}
\subsection{Pilot Test}

\begin{table}[ht]
\caption{Recognition performance leveraging different architectures of implementation.}
\label{tab:three}
\begin{minipage}{\columnwidth}
\begin{center}
\begin{tabular}{lll}
  \toprule
  \textbf{Architecture}   & \textbf{Accuracy}   & \textbf{F-Score}    \\
  Baseline(RF) + raw embeddings     & 34.4\%   & 24.5\%   \\
  Baseline(RF) + random oversampling     & 36.7\%   & 27.2\%   \\
  Baseline(RF) + SMOTE     & 45.6\%   & 37.1\%   \\
  CNN + raw embeddings     & 52.2\%   & 44.8\%   \\
  CNN + random oversampling  & \textbf{81.1\%}   & \textbf{80.0\%}   \\
  CNN + SMOTE     & 73.3\%   & 71.1\%   \\
  
  \bottomrule
\end{tabular}
\end{center}

\end{minipage}
\end{table}%

\begin{figure*}[b]
\centering
{\includegraphics[width=0.6\textwidth]{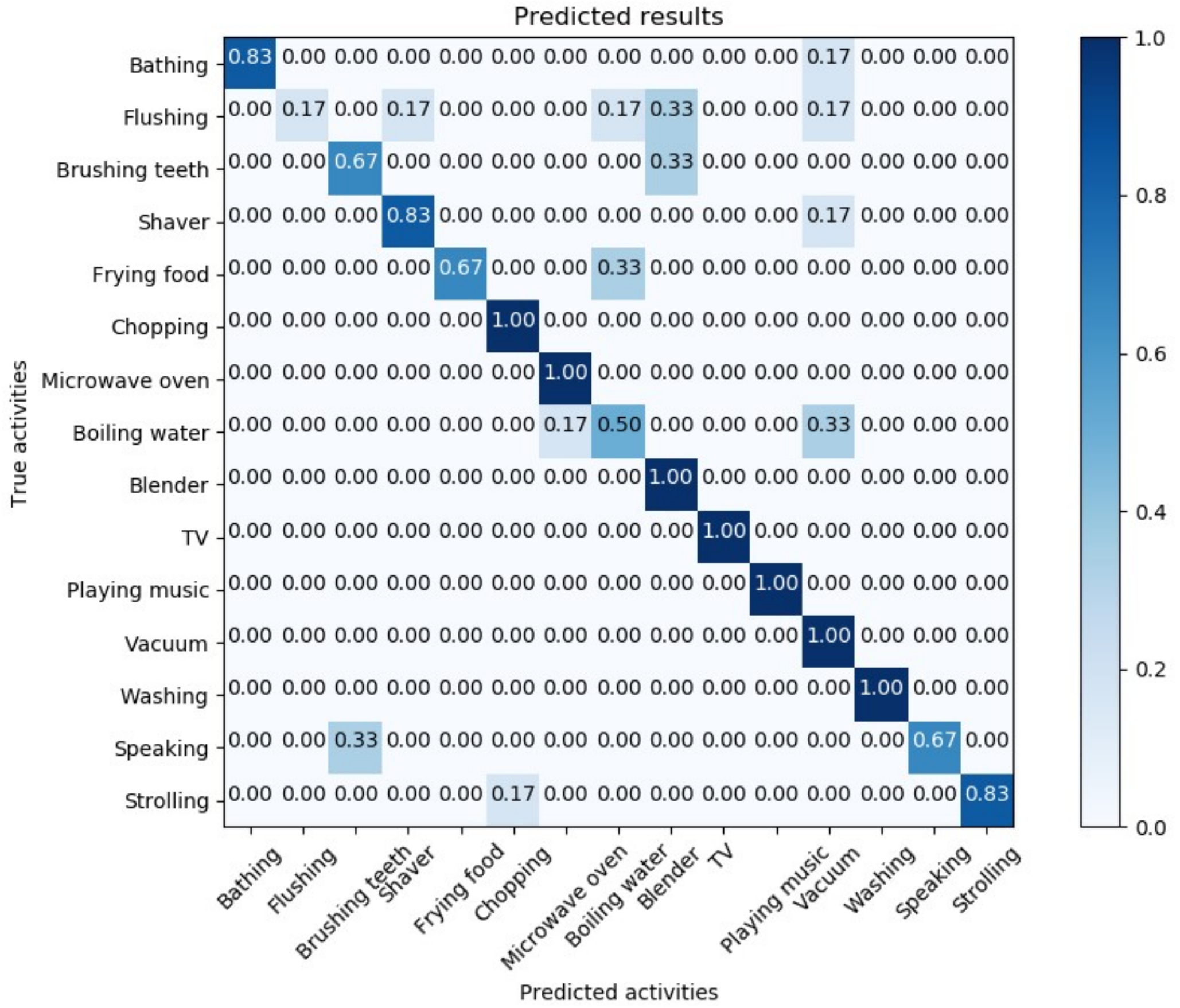}}
\label{fig_first_case}
{\includegraphics[width=0.6\textwidth]{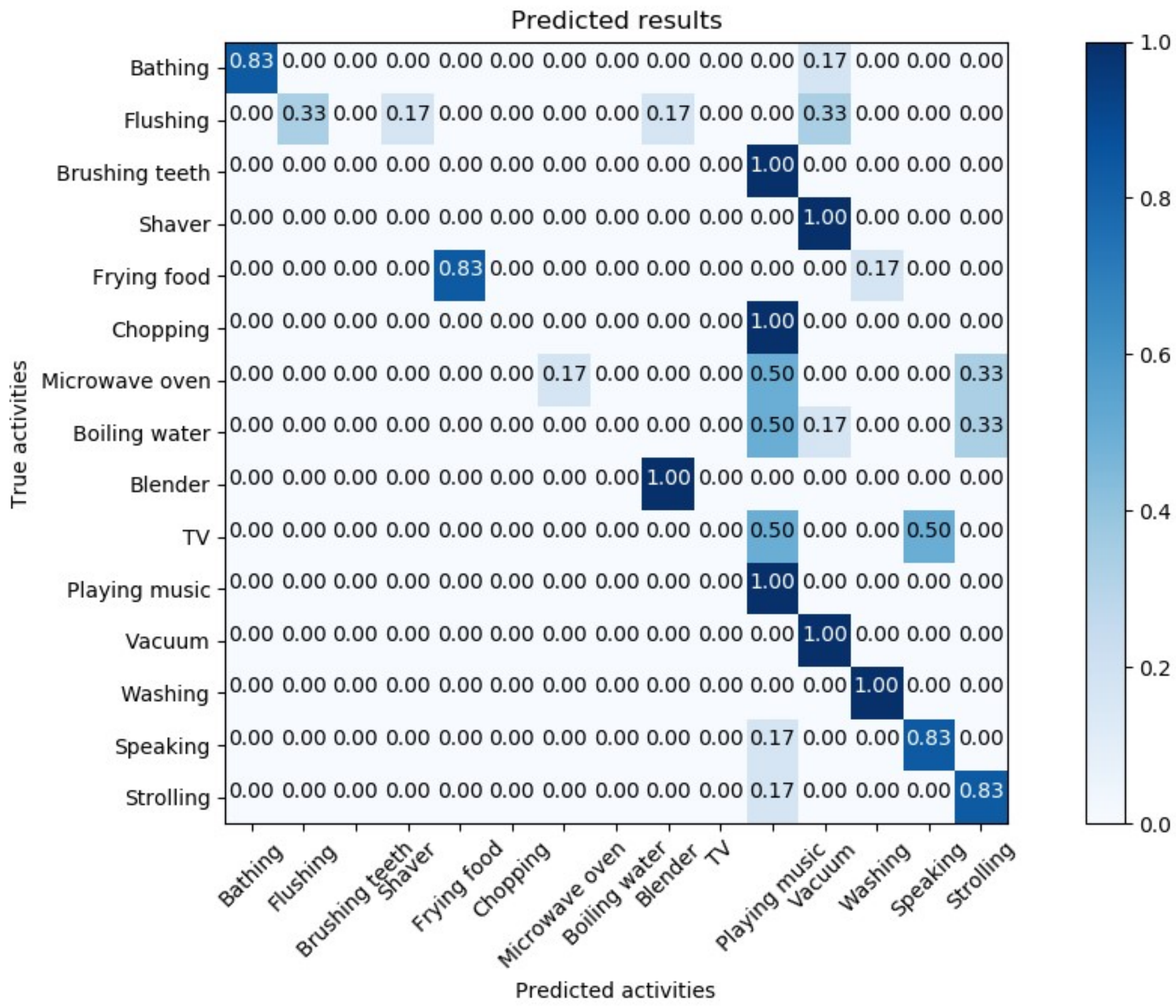}}
\label{fig_second_case}

\caption{Recognition results of the pilot study using random oversampling + CNN (top) and raw embeddings input + CNN (down). It is obvious that the performance with the oversampling process far exceeds the performance with only raw embeddings input.}
\label{fig:three}
\end{figure*}

We evaluated the feasibility of our framework based on a pilot lab study. There are two purposes to do so. Firstly, we would like to check if our proposed methodology can actually work based on real-world ambient recordings. Although the architecture had been well trained on the Audio Set data, the characteristics of the YouTube video sounds and real-world ambient sounds could possibly be different. Secondly, we would need a real-world test to determine the best combination strategy for the sampling process and the classifier. In this dedicated study, we collected sounds of the target activities in the wild by placing an off-the-shelf smart phone (Huawei P9) nearby for recording. In the pilot study, the context of the activities was well-controlled with low variability. Specifically, we tried best to exclude irrelevant environmental noise such as sounds of toilet fans or air conditioners during the collection. Also, when a target activity was performed there were no extra on-going activities. When the study began, the smart phone was placed in a natural fashion near where the activity was going to be performed. The collection was manually started when the sound of the activity could be clearly captured. Sound recording for each activity lasted for 60 seconds, and it would be stopped when the proposed time ended. This same process had been repeated for each individual activity until the collection for all 15 activities was completed.

We chose a segmentation size of 10 embedding vectors (9.6 seconds) for the dedicated study. The recognition performance was evaluated based on 3 different sampling processes (raw embeddings input/no oversampling, random oversampling, and the SMOTE). To make it clearer how the classification network performs, we also tuned and trained a random forest classifier on the same training sets as a baseline. The random forest was built using the Python Scikit-learn package \cite{scikit-learn}. We used the overall accuracy and overall F-score as the performance metrics. In binary classification, the F-score is calculated as 2 * (precision * recall) / (precision + recall) and it incorporates information for both precision and recall performance. In our study, the overall F-score across multiple classes can be calculated by finding the weighted average of F-scores of the individual labels. Table 3 shows the recognition performance based on different architectures. For convenience, the random forest is abbreviated as RF in the table. From the results we can see that the random forest without any sampling process yields the worst accuracy and F-score (34.4\% and 24.5\%). This is comparable to the dedicated study by Rossi et al. \cite{rossi2012recognizing}, where the authors trained a GMM on 4678 raw samples from the crowd-scoured Freesound dataset and obtained 38\% overall accuracy for 23 context categories. Clearly, introduction of the classification network significantly improves the recognition performance, especially if combining with the oversampling processes. The combination of random oversampling and our classification network yields the best performance (81.1\% overall accuracy and 80.0\% overall F-score). Generally, classifiers with oversampling outperform those without one. Fig.3 shows in details the performance of individual classes with and without oversampling, and the entries have been normalized for each class. As we can see, classification network input with raw embeddings overfits to some of the majority classes such as 'playing music' and 'strolling'. Network input with the random oversampled embeddings, on the contrary, yields equally promising results to most classes. The worst class for the top-1 architecture was 'flushing toilet' with only 17\% class accuracy. This is probably because the segmentation length was too long to the flushing activity and too much irrelevant information was captured within the segments.

\begin{figure}[ht]
  \includegraphics[height=0.4\textwidth]{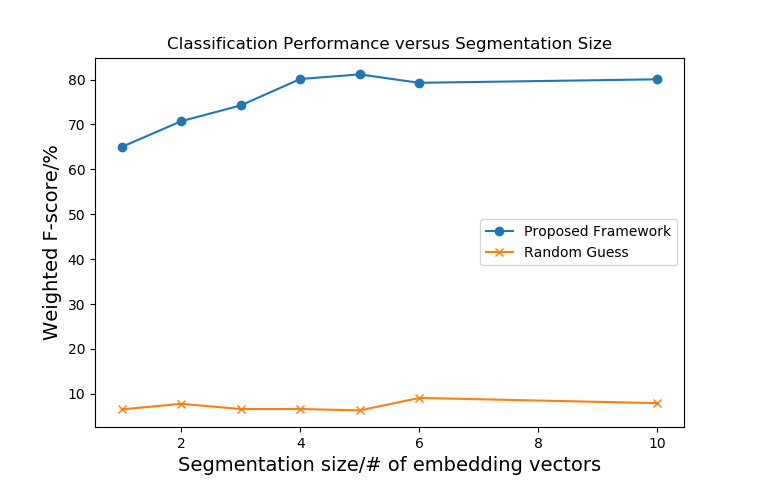}
  \caption{F-score performance with different segmentation size. The performance was worst when no segmentation process was applied. With increment of the segments size, the F-score significantly increased and remained stable around 80\%. The random guess levels were around 7\%.}
  \label{fig:four}
\end{figure}

To determine how the segmentation process can affect the classification performance, we compared the overall F-score under different size of embedding segmentation. The comparison is shown in Fig.4. As reference, we also plotted the random guess levels (around 7\%). From the figure, we can see that the performance was the worst when no segmentation process was introduced (i.e. 1 embedding vector each segment), with an F-score of only 65\%. By applying a bigger segment size, the F-score value significantly increased to over 80\%. In addition, we can see that the unit segmentation length of 5 embedding vectors has already enabled the instances to capture enough information for the classification. Further enlarging the size of segmentation no longer improve the overall recognition performance.

\subsection{More Discussions towards Domain Shifts}

From the perspective of transfer learning, our framework is actually a domain adaptation process where we try to find a mapping between the source Youtube soundtracks and the real-world recordings. Generally speaking, audio features from such on-line videos can be very different from those of the real-world collections for activity recognition. Interestingly, our classification network only yielded 53\% validation and training accuracy on the random oversampled Audio Set embeddings. But the performance of our top-1 scheme reached to over 80\% on the ambient recordings. Besides, we have noticed that the validation performance on the Audio Set data could have been further improved by adopting deeper layers. However, increasing the depth of the model would no longer help to improve the performance on real-world data (it might even harm the performance). A possible reason is that ambient sounds from the real world (especially in home settings) can generally be of less complexity and be more 'linear separable' than those on the YouTube videos. In other words, a model fit too much on the Audio Set data can probably becomes over-fitting to the sound recordings from our home settings.

%% file: Free_test.tex
\section{In-the-wild Tests}

\subsection{Test Design}

By the pilot test, we verified the feasibility of the proposed framework and determined the appropriate combination of the oversampling and segmentation strategies with the proposed networks. To generalize the study in more natural settings, we then implemented in-the-wild scripted tests based on 14 human subjects in their actual home environment. In the previous feasibility study, we made several assumptions for the test environment. Firstly, there was little irrelevant environmental noise such as noise of common home appliances during the collection process. The audio samples were recorded by a smart phone nearby with almost no artificial or ambient disturbance during the processes. Secondly, the start and end points of the collection were also carefully selected to ensure high quality recordings. Thirdly, there were almost no overlaps and co-occurrence among the activities. In other words, individual collections were ensured to be strictly mutual exclusive. However, in real-world settings such assumptions can always be broken. For example, human artifacts such as sounds from roommates and ambient noise from air conditioners or refrigerators are almost inevitable in our home. Also, people tend to perform activities in a more continuous way and it is quite reluctant if the framework always requires a pause. Hence, we are interested to see how the proposed architecture performs under such more natural circumstances. 

The real-world tests were performed based on a scripted scenario. A key advantage of the scripted tests is that the procedure of following the script can simulate the continuous process of human activities just as in natural home settings. All target activities were listed in advance in the form of instructions such as "First head to the bathroom, wash your hands and face" or "After juice prepared, please warm some food using the microwave oven". Each human subject then simply followed the instructions on a paper and freely perform the activities. We adopted the same off-the-shelf device (Huawei P9) in the collection. The smart phone was carried on the subjects' arms with a wristband so that the they could perform the activities without paying attention to the collection process. During the whole collection, an expert (one of the authors of the paper) followed the subjects while they were performing the activities but would keep a distance (e.g. waiting outside the room while the subject was performing room cleaning) to allow sufficient freedom for the subjects. The key roles of the expert were to answer questions by the subjects during the test and to label the time stamps of the target activities by using a timer started simultaneously with the recording phone. To avoid subjective bias, the tested volunteers had not been told the full purpose of the experiment until the whole collection was completed. All participants of the study were required to signed an IRB protocol form before the tests.

To incorporate variability factors in the tests, the expert would occasionally introduce a small amount of free chatting during some of the activities such as watching TV, frying or strolling. To simulate the concurrence of activities, the subjects were allowed to perform some activities simultaneously such as short washing during the frying work. In addition, all 14 tests were performed in volunteers' own home and they were allowed to leave some household appliances such as the air conditioners or refrigerator compressor working as normal. To further follow their normal modality, they were encouraged to use their own devices or tools (e.g. their own vacuum cleaner, kitchen and toilet appliances) for the collection.

In our script, most activities were required to be just performed once and the length was determined freely by the participants. We prepared some bacon, cucumbers or carrots in advance for activities 'frying food', 'chopping food' and 'squeezing juice'. Given the diversity of television programs, the participants were asked to watch for 5 different channels with around 30 seconds each for the class 'watching TV'. For class 'enjoying music', the subjects were specified to play their own piano or listen to relevant types of musics such as piano solo or symphonies chosen by themselves. Besides, class 'shavering' was waived for female subjects.

\subsection{Results and Discussions}

In total we were able to obtain 32105 seconds (535 minutes) of audio data from 14 subjects (7 males and 7 females). Based on the labeled time stamps, we manually segmented the target activity data from the raw recordings. Overall, we identified that roughly 12078 seconds (201 minutes) of the clips were target-related, accounting for 37.6\% of the total. The resulting sparsity is comparable to audio-based activity recognition in practice as not all home-related activities can generate specific sound features and audio-based frameworks are not suitable for them. We then applied the best architecture of the proposed framework (classification network with random oversampling) to evaluate the results. The unit segmentation length was set as 10 embedding vectors (9.6 seconds).

\begin{figure}[ht]
  \includegraphics[height=0.4\textwidth]{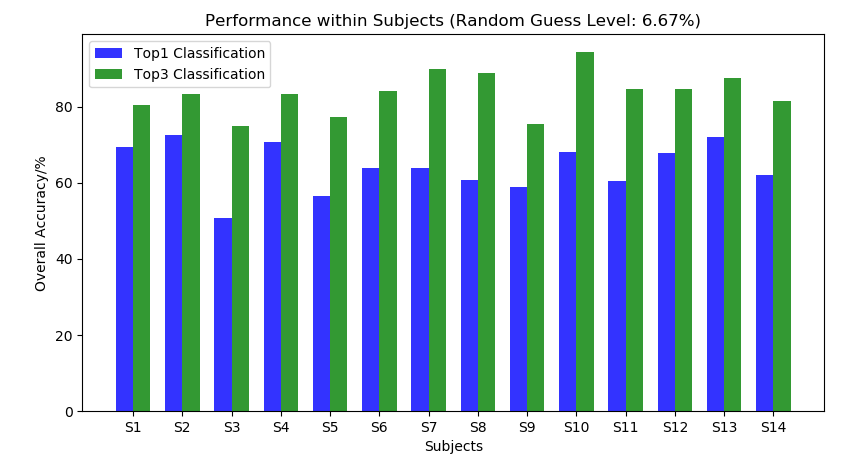}
  \caption{Classification performance within subjects. The averaged top-1 and top-3 accuracies are 64.16\% and 83.59\% respectively for all subjects.}
  \label{fig:five}
\end{figure}

The test results were first examined based on each individual participant. Fig.5 shows the overall classification performance within single subjects. Because of the high inequality of segment length among the activities, we adopted the overall weighted average as the performance metric. In other words, for a given subject, the contribution of each tested instance to the overall accuracy is inversely proportional to the amount of tested data within that corresponding activity class. By weighting the instances, each activity class within the subject can then contribute equally to the overall performance. In our studies, the averaged top-1 classification accuracy was 64.16\% for all tested subjects. In addition to the top-1 classification, we also evaluated the overall performance using a top-3 classification scenario given the co-occurrence of activities and the variability during the tests. In the top-3 classification, predicted labels with the top 3 highest probability are considered as the final predictions, and a true positive can be counted if any of the 3 labels match the ground truth. It incorporates the variants of predictions due to possible similarity of sound features or concurrence of the actual activities. From the figure we can see that the top-3 performance was much better than the top-1 scenario, with an averaged accuracy of 83.59\% for all 14 subjects. 

\begin{figure}[b]
  \includegraphics[height=0.4\textwidth]{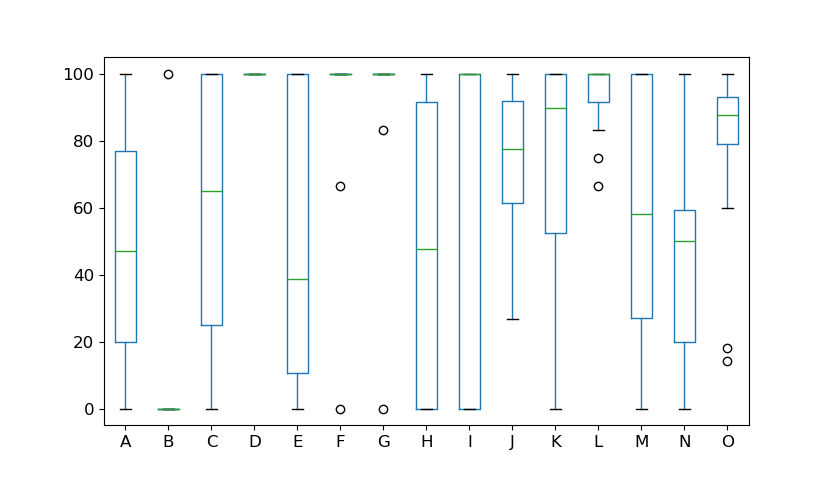}
  \caption{Top-1 classification performance for individual activity classes across the subjects (A:Bathing/Showering; B:Flushing; C:Brushing Teeth; D:Doing Shaver; E:Frying; F:Chopping; G:Microwave Oven; H:Boiling; I:Squeezing Juice; J:Watching TV; K:Playing Music; L:Floor Cleaning; M:Washing; N:Chatting; O:Strolling)}
  \label{fig:six}
\end{figure}

\begin{figure}[ht]
  \includegraphics[height=0.4\textwidth]{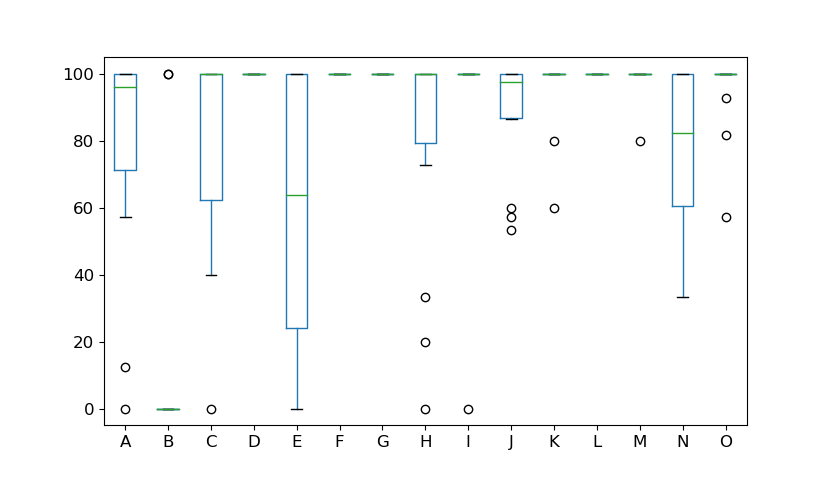}
  \caption{Top-3 classification performance for individual activity classes across the subjects (A:Bathing/Showering; B:Flushing; C:Brushing Teeth; D:Doing Shaver; E:Frying; F:Chopping; G:Microwave Oven; H:Boiling; I:Squeezing Juice; J:Watching TV; K:Playing Music; L:Floor Cleaning; M:Washing; N:Chatting; O:Strolling)}
  \label{fig:seven}
\end{figure}

To evaluate the performance of individual activity classes, we also summarized the class accuracies across all tested subjects. We calculated the average values for both the top-1 and top-3 classification, and Fig.6 and Fig.7 present the statistics for both settings. Instead of directly applying confusion matrices, we adopted a similar weighted approach for the analysis. That is, tested instances from each subject were assigned with weight that was inversely proportional to the amount of data within them. This enables samples from different subjects and different tested environment with varying data size to contribute equally to the overall performance of the target classes. In addition, the figures also indicate the deviations of the class accuracies away from the mean. A smaller deviation represents a more stable performance of the predictions and further implies a stronger robustness of the framework towards variants in the actual tests. As indicated from the figures, 'shavering', 'chopping food' and 'using microwave oven' showed the best performance with almost 100\% averaged class accuracy and almost zero deviation. Class 'floor cleaning' was also of satisfactory results due to its clear and unique sound features. On the contrary, however, most of the flushing activities were misclassified by the framework. It is probably because the process of pumping was too short given the segmentation length and the sounds of water flushing can largely overlap with those of the washing or frying activities. Also in the figures, some activities such as 'frying food', 'boiling water', 'squeezing juice' and 'brushing teeth' are of high deviations from their average. It is reasonable because the modalities of cooking and boiling can vary in practice depending on the choice of the kettles and cooking tools or the variant cooking styles among the participants. The performance of kitchen activities was also affected by usage of hoods by some of the participants. The brushing activity could mainly be affected by the noise of toilet fans. Especially, we noticed that our framework failed to recognize almost all brushing activities with electric toothbrush possibly due to the lack of relevant training samples in the Audio Set. If comparing the results in both figures, we can also see that the performance of most activities increased significantly from the top-1 scenarios to the top-3 scenarios, reaching to nearly 100\% mean accuracy with much smaller deviations. This implies the existence of activity co-occurrence and overlaps of acoustic features among distinct activities such as simultaneous chatting with outdoor strolling or a music show on TV, which are also commonly seen in the natural home settings.

Because of the difference in terms of evaluation metrics and test conditions, it was challenging to directly compare the performance across the related work. As reference, Rossi et al. \cite{rossi2012recognizing} combined a semi-supervised or manual filtering of outliers with the Gaussian Mixture Model (GMM) to classify 23 acoustic contexts. They extracted the MFCC features from the Freesound dataset with the sequence length of 30 seconds for training. The best top-1 classification and top-3 classification performance were 57\% and 80\% respectively only if with manual filtering of the outliers. Hershey et al. \cite{hershey2017cnn} trained two fully connected networks with and without the embedding extraction process to classify the Audio Set \cite{gemmeke2017audio} categories. They adopted the mean Average Precision (mAP) as the performance metric and obtained the best mAP of 0.31 only if taking the embeddings as input. Kong et al. \cite{kong2017audio} completed a similar test using an attention model from a probability perspective, achieving mAP of 0.327 and AUC of 0.965. The state of the art by Laput at al. \cite{laput2018ubicoustics} reported the classification performance from several perspectives. Their best model achieved 80.4\% overall accuracy for 30 context classes recorded in the wild, but the framework relied on a mixed process of audio augmentation and combination of sound effect libraries for training. If purely using the online video sounds (i.e. the Audio Set \cite{gemmeke2017audio} data), their framework yielded an overall accuracy of 69.5\% when check-pointed on the test set and 41.7\% when tested directly on the real-world sounds. Correspondingly, our framework was not developed based on any feature augmentation and semi-supervised learning processes. The overall classification accuracy of our model was 81.1\% for 15 activity classes in the lab study. Our top-1 and top-3 performance was 64.2\% and 83.6\% respectively based on multi-subject tests of 14 participants in their actual home environment.

%% file: conclusion.tex
\section{conclusion}

The collection of ground truth user data can be time-consuming and laborious in multi-class audio learning. This paper presented a novel framework leveraging large-scaled on-line YouTube video soundtracks as the only training set to empower audio-based activity recognition. Specifically, our proposed framework aims to recognize 15 common home-related activities. Due to the tremendous size of the dataset and the highly unbalanced distribution of the training classes, our framework combined both oversampling and deep learning architectures without further needs of feature augmentation and semi-supervised learning processes. To evaluate its performance, we designed both in-lab pilot tests and in-the-wild scripted tests with multiple subjects in their home. Results showed that our proposed framework was able to achieve promising performance and robustness to the environmental variability in different test scenarios. Other design considerations including the association of activity labels and effects of embedding segmentation were also discussed in the paper.